\documentclass[sigconf] {acmart}
\AtBeginDocument{%
  }

\setcopyright{acmlicensed}
\copyrightyear{2026}
\acmYear{2026}
\setcopyright{cc}
\setcctype{by}
\acmConference[CHI '26]{Proceedings of the 2026 CHI Conference on Human Factors in Computing Systems}{April 13--17, 2026}{Barcelona, Spain}
\acmBooktitle{Proceedings of the 2026 CHI Conference on Human Factors in Computing Systems (CHI '26), April 13--17, 2026, Barcelona, Spain}
\acmDOI{}
\acmISBN{}

\newcommand{\privy}{\textsc{Privy}}

\usepackage{tabularray}
\usepackage{footnote}
\makesavenoteenv{tabular}
\makesavenoteenv{table}
\usepackage{csquotes}
\usepackage{xcolor} 
\usepackage[colorinlistoftodos, textsize=tiny, backgroundcolor=yellow!20]{todonotes}
\usepackage{graphicx}
\usepackage{colortbl}
\usepackage{multirow}

\usepackage{longtable}
\usepackage{array}
\usepackage{booktabs}

\usepackage{graphicx}
\usepackage{subcaption}
\usepackage{tabularx}
\usepackage{multirow}
\usepackage{pdfpages}
\usepackage{booktabs}
\usepackage{makecell} 
\usepackage{ccicons}

\definecolor{LLM}{HTML}{F6D6AD}
\definecolor{Template}{HTML}{E3C5F2}
\definecolor{quote}{HTML}{7B3F00}
\usepackage{caption} 

\begin{document}

\title{Promoting Critical Thinking With Domain-Specific Generative AI Provocations}


\author{Thomas Serban von Davier}
\affiliation{%
  \institution{Carnegie Mellon University}
  \city{Pittsburgh}
  \state{PA}
  \country{United States}
}
\email{tvondavi@andrew.cmu.edu}

\author{Hao-Ping (Hank) Lee}
\affiliation{%
  \institution{Carnegie Mellon University}
  \city{Pittsburgh}
  \state{PA}
  \country{United States}
}
\email{haopingl@cs.cmu.edu}

\author{Jodi Forlizzi}
\affiliation{%
  \institution{Carnegie Mellon University}
  \city{Pittsburgh}
  \state{PA}
  \country{United States}
}
\email{forlizzi@cs.cmu.edu}

\author{Sauvik Das}
\affiliation{%
  \institution{Carnegie Mellon University}
  \city{Pittsburgh}
  \state{PA}
  \country{United States}
}
\email{sauvik@cmu.edu}

\renewcommand{\shortauthors}{von Davier et al.}

\begin{abstract}
  The evidence on the effects of generative AI (GenAI) on critical thinking is mixed, with studies suggesting both potential harms and benefits depending on its implementation. Some argue that AI-driven provocations, such as questions asking for human clarification and justification, are beneficial for eliciting critical thinking. Drawing on our experience designing and evaluating two GenAI-powered tools for knowledge work, ArtBot in the domain of fine art interpretation and Privy in the domain of AI privacy, we reflect on how design decisions shape the form and effectiveness of such provocations. Our observations and user feedback suggest that domain-specific provocations, implemented through productive friction and interactions that depend on user contribution, can meaningfully support critical thinking. We present participant experiences with both prototypes and discuss how supporting critical thinking may require moving beyond static provocations toward approaches that adapt to user preferences and levels of expertise.
\end{abstract}




\maketitle

\section{Introduction}
As generative artificial intelligence (GenAI) systems continue to increase in popularity and capability, applied research has increasingly demonstrated the benefits of domain-specific models over large, general-purpose models, particularly in settings that require nuanced reasoning or contextual sensitivity \cite{liu2025harnessingcollaborativepowerlarge, hsieh2023distillingstepbystepoutperforminglarger, belcak2025smalllanguagemodelsfuture}. In light of recent discussions on the negative impact of GenAI on critical thinking, these findings suggest that specialization, rather than scale alone, can significantly improve the outcomes of human-AI interaction.

In this workshop paper, we argue that a similar principle applies to the design of GenAI systems intended to support critical thinking, particularly when such systems are framed as provocateurs or facilitators. Prior work has emphasized the importance of deliberately antagonistic or challenging AI behaviors as a core design requirement for tools for thought \cite{sarkar2024copilotautopilotgenerativeais,cai2024antagonisticai}. This line of research argues that when AI systems introduce friction by questioning assumptions, surfacing alternatives, or resisting user intent, they prompt users to pause, reflect, and engage in deeper critical thinking during interaction.

Building on this literature, we advance the position that provocations embedded in GenAI systems must themselves be domain-specific to effectively support critical thinking. Rather than relying on general-purpose challenges, we contend that provocations grounded in domain knowledge, norms, and established frameworks are more likely to be meaningful, interpretable, and actionable for users.

To support this argument, we draw on our experience designing and evaluating two GenAI systems presented in recent CHI papers, each developed explicitly to support critical thinking within a distinct domain. The first system acts as a conversational companion for digital collections of fine art, where provocations are informed by art history and educational curricula to encourage interpretive reflection \cite{vonDavier2025ArtBot}. The second system is a structured, AI-assisted whiteboard for practitioners to identify privacy risks and mitigations, with provocations derived from established privacy taxonomies and frameworks. Although both systems adopt a provocateur–facilitator framework, the form and content of their challenges are tightly coupled to their respective domains \cite{lee2026privyenvisioningmitigatingprivacy}.

In this paper, we present an overview of both systems, including how domain knowledge informed the design of provocations and how each system was evaluated with users. From these cases, we surface comparative learnings about how domain-specific provocations are received by different audiences, as well as design insights into how framing GenAI as a provocateur or facilitator influences engagement and interpretation.

Ultimately, this workshop contribution offers preliminary recommendations for framing GenAI systems as effective tools for domain-specific critical thinking. We argue that while domain specificity strengthens the impact of provocations, user preferences and expectations continue to shape how such systems are perceived and used. We present this work to invite discussion on how to balance domain grounding, provocation, and user agency in future tools for thought.
\section{Background}
While industry, government, and the research world explore various paths towards improving AI and building more AI-enabled systems, there has been a growing field of work questioning whether the current implementation of automation-prioritizing chatbots is truly the most effective form of tooling. A large survey and mixed-methods study by Microsoft researchers found that many knowledge workers report lower participation and lower engagement with critical-thinking steps when they rely on GenAI for writing \cite{Lee2025Thinking}. Similarly, a widely reported neurocognitive study from the MIT Media Lab used EEG to compare people writing essays with ChatGPT, with search, or unaided \cite{kosmyna2025brainchatgptaccumulationcognitive}. The LLM-assisted writers showed reduced neural markers of executive control, memory encoding, and creativity in the tasks used by the study, a signal that heavy reliance on AI for generative work can reduce the cognitive activation behind learning and original composition. This reveals a ``payoff'' problem that some interpret as evidence that human productivity gains from AI are neither automatic nor uniformly distributed. Taken together, these strands suggest that (a) AI can lower cognitive effort and measurable engagement in specific tasks, (b) reported productivity gains do not always translate into better-quality outcomes, and (c) design choices matter if we want AI to augment rather than atrophy human critical thinking.

Many researchers, including our colleagues interested in the Tools for Thought workshop, argue that GenAI systems should scaffold, provoke, and support deliberation instead of simply producing answers. The principle behind viewing LLM interactions this way is the same principle that guides peer review, testing \& evaluation, iteration and the Socratic method: that an idea or plan is only considered high quality if it has been challenged and verified. Researchers have proposed interventions (e.g., ``provocations'' \cite{sarkar2024copilotautopilotgenerativeais} or ``antagonisms'' \cite{cai2024antagonisticai}) that deliberately surface critiques or alternatives to model outputs; their work shows such micro-interventions can increase metacognitive activity and user scrutiny in shortlisting and knowledge tasks. This research philosophy has contributed to Park and Kulkarni’s thinking assistant \cite{park2024thinkingassistantsllmbasedconversational} and Liu et al.’s Thoughtful AI \cite{liu2025interactingthoughtfulai}. 

Ye et al. explore how AI could be engineered to ask better questions and support domain-specific inquiry rather than provide ready-made conclusions \cite{ye2024languagemodelscriticalthinking}. A focus on domain-specific inquiry led to our ArtBot paper \cite{vonDavier2025ArtBot}. It is an example of a Socratic-style companion that guides users through analysis by prompting reflection and layered questioning, illustrating how dialogic agents can foster analytic practice in nontechnical domains. Other tools like FarSight \cite{wang2024FarSight} and Privy \cite{lee2026privyenvisioningmitigatingprivacy} demonstrate in-situ interfaces that scaffold human-AI reasoning within established structures. These tools show how embedding reflective checks directly into development and authoring environments can shift outcomes away from ``copy-paste'' convenience and towards accountable, deliberate practice.
\section{Case Studies}
Our prior work introduced two GenAI systems designed to support distinct forms of knowledge work through domain-specific provocations: ArtBot, a conversational companion for fine art interpretation, and Privy, a structured ideation tool for identifying and mitigating AI privacy risks. Both systems were developed and evaluated in recent CHI papers, in which their task-specific effectiveness was assessed through controlled experiments. In this workshop paper, we focus not on comparative performance outcomes, but on the design methodologies and interaction paradigms that distinguish the two systems and inform their role as tools for critical thinking.
Both systems were intentionally designed to be compared against non–AI-powered alternatives and were evaluated with participants in controlled settings (ArtBot: \textit{n} = 13; Privy: \textit{n} = 12\footnote{The original paper presents two versions of Privy \cite{lee2026privyenvisioningmitigatingprivacy} --- one incorporating LLM-powered features and one without --- each evaluated with 12 practitioners.}). In addition to addressing their primary research questions, these studies generated rich observational data on how users responded to different forms of AI provocation and facilitation. Below, we outline the core design decisions underpinning each system, with particular attention to how domain knowledge shaped the structure and presentation of provocations.

\begin{table*}[t]
\caption{This table provides an overview on the theoretical grounding underlying both systems and which specific domains each one operates under.}
\resizebox{\textwidth}{!}{%
\begin{tabular}{|l|l|l|l|}
\hline
\textbf{Case Study} & \textbf{Domain}             & \textbf{Theoretical Grounding}           & \textbf{Model Type}                     \\ \hline
ArtBot     & Art Interpretation & Dialogic Education \cite{Skidmore2016}             & Local, open-weight models \cite{Llama3_1}      \\ \hline
Privy      & AI Privacy Reports & Framework Grounded Provocations \cite{das2022security} & Cloud-host, proprietary models \cite{wei2022chain} \\ \hline
\end{tabular}%
}
\end{table*}

\subsection{ArtBot: Domain-Grounded Provocation for Art Interpretation}
\begin{figure}[h]
  \centering
  \includegraphics[width=0.85\textwidth]{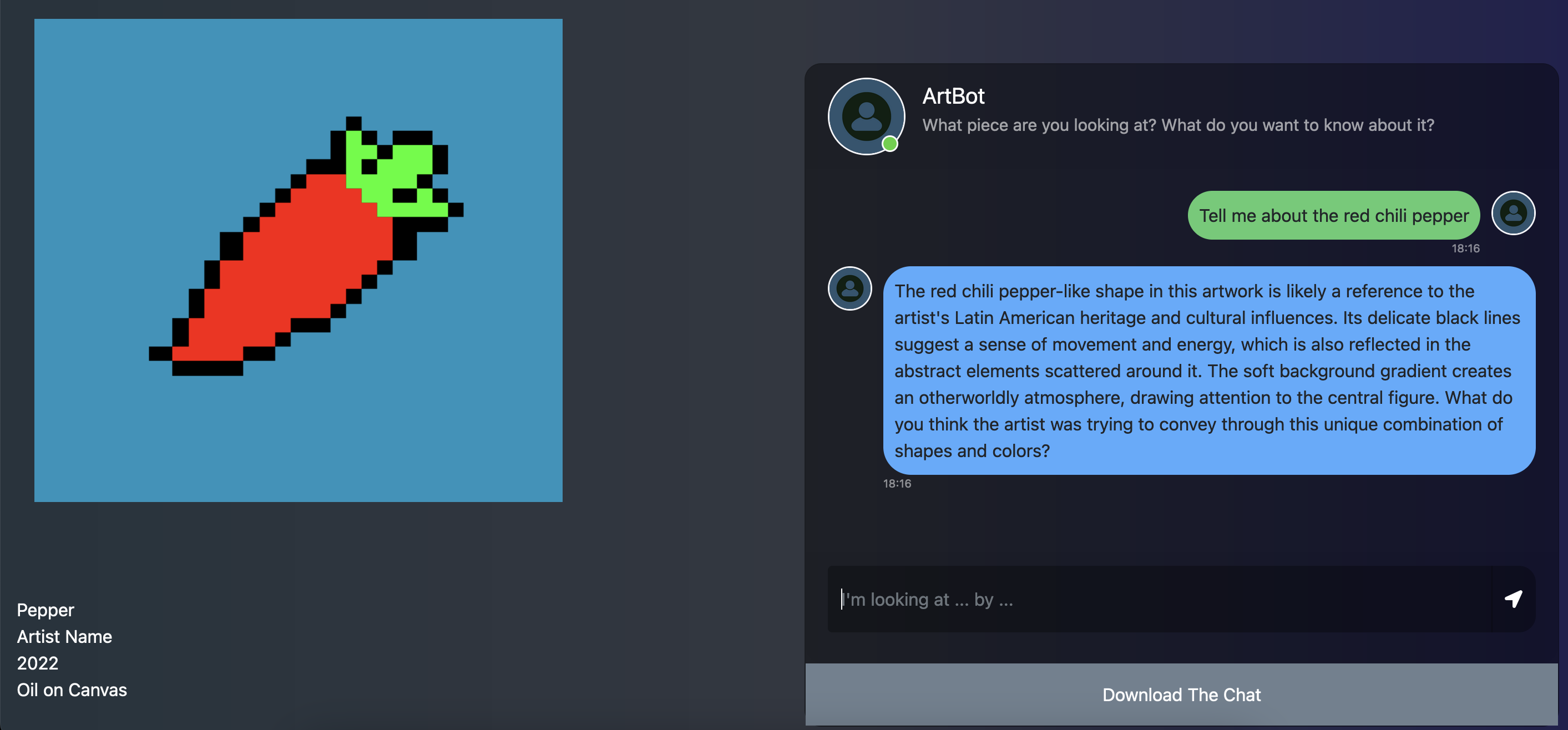}
  \vspace{-2mm}
  \caption{\textbf{ArtBot} is an LLM-powered tool that challenges participants with questions to encourage them to share their own interpretations of the artwork in the shared workspace.}
  \label{fig:artbot}
\end{figure}

ArtBot (Figure \ref{fig:artbot}) was developed to support the engagement with digital fine art collections through an interactive, large language model–augmented interface \cite{vonDavier2025ArtBot}. The system was implemented using locally hosted Llama 3 models \cite{Llama3_1} combined with RAG (Retrieval Augmented Generation) \cite{Lewis2020RAG}, allowing access to a curated corpus of art-historical metadata, curatorial texts, and educational materials.

The interaction paradigm of ArtBot draws inspiration from Socratic tutoring, an instructional approach in which understanding is developed through targeted questioning rather than direct explanation. This approach aligns with dialogic education practices, where learning emerges through a dialogue of provocations and reflective responses \cite{Skidmore2016}. In ArtBot, these practices were operationalized through prompts that challenge users to articulate interpretations and reconsider assumptions about an artwork. Some examples of the interaction include the GenAI system asking the participant whether their interpretation of the artwork changes if they know it was made during a time of revolution, and the participant then responding and discussing whether their interpretation changes and how.

Users interact with ArtBot while viewing an artwork image, engaging in a conversational exchange intended to deepen interpretation rather than deliver authoritative explanations. During the evaluation, participants were asked to provide short written reflections after interacting with each artwork, which served as the basis for assessing interpretive engagement.

\subsection{Privy: Structured Provocation for AI Privacy Planning}
\begin{figure}[h]
  \centering
  \includegraphics[width=0.95\textwidth, trim=0 0 0 2, clip]{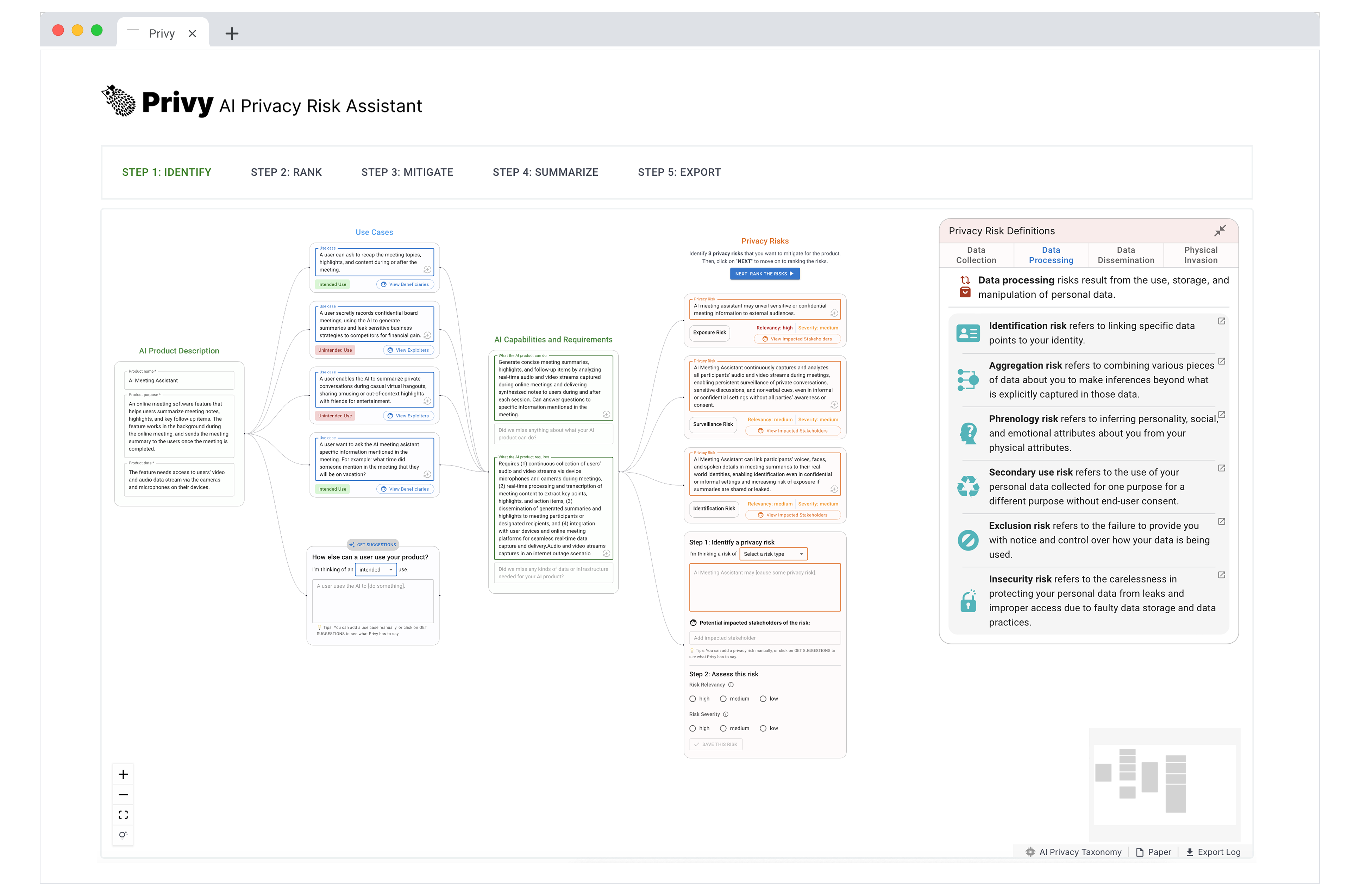}
  \vspace{-5mm}
  \caption{\textbf{\privy{}} is an LLM-powered tool that guides practitioners through structured privacy impact assessments to: (i) identify relevant risks in novel AI product concepts, and (ii) propose appropriate mitigations.}
  \label{fig:privy}
\end{figure}

Privy (Figure \ref{fig:privy}) was designed to support AI practitioners during the early design and planning phases of AI system development, with a focus on identifying and mitigating privacy risks \cite{lee2026privyenvisioningmitigatingprivacy}. Privy leverages a structured, branching-tree workflow embedded within a whiteboard-style interface. This structure guides users through a sequence of decisions and reflections commonly encountered in AI system design.

The system is augmented by a GenAI backend (implemented using GPT-4.1) and is informed by system prompts grounded in established AI privacy taxonomies, including Lee et al. and Das et al.’s frameworks \cite{Lee2024Taxonomy, das2022security}. These domain-specific prompts (e.g., design frictions that require users to assess the relevance and severity of risks, or to reflect on the effectiveness of proposed mitigations) enable Privy to challenge practitioners by surfacing potential risks, highlighting unintended use cases, and facilitating planning for privacy mitigation best practices --- e.g., \textit{How can you design this feature to encourage users to regularly review and update their sharing settings so they stay in control of how their social network data is used?} 

As users progress through the Privy workflow, they iteratively document privacy risks and associate each with tailored mitigation strategies. The interaction culminates in a structured artifact, a design document summarizing identified risks and proposed mitigations. Evaluation focused on the quality and completeness of these artifacts, which were assessed by privacy experts.

\section{Supporting Critical Thinking Through Challenging Provocations}
Although ArtBot and Privy differ substantially in domain, audience, and interaction style, both systems were intentionally designed to position GenAI as a provocateur and facilitator rather than an answer engine. In each case, domain knowledge plays a central role in shaping how provocations are formulated, when they are introduced, and how users are encouraged to respond.

Both ArtBot and Privy intentionally incorporated \textbf{design friction} into their interactions by resisting the impulse to provide direct answers. In ArtBot, this friction took the form of a Socratic interaction style that prompted users to articulate their own interpretations of an artwork instead of presenting curator-authored wall text. Similarly, within the mitigation stage of the Privy workflow, recommended privacy mitigations were framed as questions rather than as a numbered list of solutions.

Across both systems, participants frequently responded to these provocations by pausing to consider their own perspectives before continuing. In some cases, this pause introduced mild frustration, particularly when users expected the system to provide more immediate or authoritative information. In other cases, participants described the questioning as engaging or surprising, noting that it surfaced considerations they had not previously explored. These reactions suggest that carefully designed friction can prompt reflection, though it may also challenge user expectations shaped by more answer-oriented AI systems.

A second shared design decision was the use of \textbf{user-created content gates}, where progression through the system required participants to contribute their own thoughts before receiving additional AI support. From a methodological perspective, this design explicitly operationalized a human-in-the-loop approach: users were required to actively engage and externalize their reasoning before the system responded.

In both ArtBot and Privy, these gates were supported by lightweight instructional text within input fields to help users structure their responses. We observed that the quality and specificity of user input often shaped the relevance and usefulness of subsequent system output. This design choice reinforced the role of the GenAI system as a facilitator of thinking rather than a generator of standalone insight, while also making user effort a visible part of the interaction.

A third design decision involved \textbf{system framing}: i.e., how each system was presented to participants. Both ArtBot and Privy were presented as possessing relevant domain knowledge and as tools intended to support human thinking within their respective domains. This framing was generally effective, but it also elicited varied reactions depending on participants’ domain expertise.

Participants with greater subject-matter familiarity were more likely to challenge the system’s suggestions, tone, or assumptions, at times expressing disagreement or skepticism. Other participants, particularly those with less experience in the domain, described the systems as informative and well-grounded. These differing responses suggest that framing GenAI as a facilitator interacts with users’ prior knowledge, influencing whether provocations are perceived as helpful, restrictive, or misaligned.

Across both systems, we observe that altering the structure of human–AI interaction through questioning strategies, workflow constraints, and domain-grounded prompts can meaningfully influence how users externalize, refine, and develop their own thinking. These shared design principles form the basis for the comparative reflections we bring to the workshop discussion.
\section{Discussion}
Across both ArtBot and Privy, domain-specific provocations appeared to support a wider range of ideas and responses than generic prompts would likely have produced. Participants drew on domain-relevant concepts, vocabularies, and concerns when responding to system questions, resulting in outputs that were better aligned with the goals of each task. Based on our observations and participant feedback, we speculate that if these systems relied on more general provocations (like simply asking ``Why?'' or asking for justification), the reflective gains observed in prior evaluations would have been diminished.

At the same time, the variability in user reactions points to an additional human factor that warrants further attention. Individual expectations, expertise, and tolerance for friction all shape how provocations are received. This suggests that while domain specificity strengthens GenAI’s role in supporting critical thinking, it does not fully determine user experience, and thereby highlights the need for adaptable or customizable provocation strategies in future systems.

One recurring pattern across both systems concerned participants’ underlying assumptions about what GenAI should do. Some participants approached the systems with a strong expectation that GenAI functions primarily as an automation tool. In ArtBot, these users expressed a desire for authoritative interpretations of artworks, preferring curator-approved explanations over dialogic exploration. Similarly, in Privy, such participants expected the system to automatically generate a completed end-to-end privacy review or privacy impact assessment (PIA) based on a brief system description.

From this perspective, the inclusion of GenAI raised a fundamental question: if the system already has access to relevant knowledge, how can users meaningfully engage in reflection or decision-making, if at all? This tension surfaced resistance to provocations designed to slow interaction or demand user input. As part of our workshop contribution, we aim to discuss how GenAI tools designers and developers might either accommodate or intentionally challenge entrenched views of GenAI as automation, drawing on prior work that explores how systems can surface, negotiate, or reframe users’ mental models at the point of interaction \cite{Wang2025MentalModels, Rezwana2022UserPerception}. The results of our case studies provide evidence supporting the interaction-automation conflict outlined by Wiberg \& Berqvist \cite{Wiberg2023Automation} which is essentially a paradox in human-AI interaction \cite{Salma2025Paradox} where the need to design meaningful interactions encounters AI’s increasing ability to automate processes.

An opposing, but equally consequential, pattern emerged among participants who expressed skepticism about GenAI’s capacity to contribute meaningfully to critical thinking. These users often referenced existing literature or public discourse critiquing GenAI’s limitations and were inclined to postpone engagement with AI-generated content for as long as possible. Several explicitly stated a preference for producing their own ideas before consulting the system.

While this behavior can be interpreted as desirable by signaling an awareness of the value of independent thinking, it also introduced new challenges. In some cases, participants underutilized features intended to expand perspective, surface blind spots, or support ideation. Prior explainable AI (XAI) work suggests that making system capabilities, limitations, and intent legible to users can help mitigate such resistance \cite{Laato2022ExplainAI}. We see an opportunity to explore how GenAI systems might better signal their role as bounded, reflective collaborators rather than as authoritative or generative replacements. We are inspired by work looking towards creating effective documentation, tutorials, and model cards \cite{piorkowski2020evaluatingelicitinghighqualitydocumentation, Crisan2022ModelCard}. 

A third tension emerged around participants’ subject-matter familiarity. Users with substantial domain expertise occasionally perceived system provocations as condescending or redundant, particularly when the system challenged assumptions they already considered well understood. Conversely, novice users sometimes interpreted the same systems as possessing expert-level authority, potentially over-weighting their suggestions.

These reactions underscore the risk of a mismatch between system tone and user expertise and perception of the GenAI system \cite{AMROLLAHI2026103031}. They also suggest that simply embedding provocation (whether general or domain-specific) is insufficient. Instead, effective support for critical thinking may require systems that are responsive to signals of user expertise, confidence, or intent, adapting their provocations accordingly.

Taken together, these reflections point to a broader design challenge: critical thinking is not only shaped by what provocations are presented, but by what role the user believes the system plays, and who they believe themselves to be in relation to it. Domain specificity strengthens the relevance and interpretability of provocations, but individual differences in expectation, expertise, and trust continue to shape outcomes.

We bring these observations to the workshop to invite discussion around how domain-specific GenAI systems might better balance provocation, adaptability, and user agency. In particular, we are interested in exploring strategies for customizing or negotiating provocation styles in response to human-centered factors. In doing so we move beyond static designs toward more responsive tools for thought.
\section{Conclusion}
Through the design and evaluation of two distinct GenAI prototypes, we developed insights into how domain-specific design decisions and implementation strategies can support critical thinking in practice. Our experiences suggest that provocations grounded in domain knowledge can more effectively support users’ knowledge work. Specifically because they are designed to introduce productive friction, invite user interaction, and frame the system as a knowledgeable facilitator. At the same time, our work highlights the role of user preferences and expectations, underscoring the need for adaptive and flexible implementations of GenAI systems. We present these cases to the Tools for Thought workshop as concrete, evidence-informed examples, and to invite discussion around how domain grounding, provocation, and user agency might be balanced in future GenAI tools for critical thinking.

\begin{acks}
We want to thank our collaborators on the original ArtBot and Privy papers, many people worked hard to bring these systems to life. We also want to thank our participants for their time and feedback.
\end{acks}

\bibliographystyle{ACM-Reference-Format}

\begin{thebibliography}{27}


\ifx \showCODEN    \undefined \def \showCODEN     #1{\unskip}     \fi
\ifx \showISBNx    \undefined \def \showISBNx     #1{\unskip}     \fi
\ifx \showISBNxiii \undefined \def \showISBNxiii  #1{\unskip}     \fi
\ifx \showISSN     \undefined \def \showISSN      #1{\unskip}     \fi
\ifx \showLCCN     \undefined \def \showLCCN      #1{\unskip}     \fi
\ifx \shownote     \undefined \def \shownote      #1{#1}          \fi
\ifx \showarticletitle \undefined \def \showarticletitle #1{#1}   \fi
\ifx \showURL      \undefined \def \showURL       {\relax}        \fi
\providecommand\bibfield[2]{#2}
\providecommand\bibinfo[2]{#2}
\providecommand\natexlab[1]{#1}
\providecommand\showeprint[2][]{arXiv:#2}

\bibitem[Amrollahi et~al\mbox{.}(2026)]%
        {AMROLLAHI2026103031}
\bibfield{author}{\bibinfo{person}{Alireza Amrollahi}, \bibinfo{person}{Jiaqi Yang}, \bibinfo{person}{Syed Muhammad~Fazal e Hasan}, {and} \bibinfo{person}{Basma Badreddine}.} \bibinfo{year}{2026}\natexlab{}.
\newblock \showarticletitle{Knowledge workers’ trust and reception of generative AI’s advice in complex tasks}.
\newblock \bibinfo{journal}{\emph{International Journal of Information Management}}  \bibinfo{volume}{88} (\bibinfo{year}{2026}), \bibinfo{pages}{103031}.
\newblock
\showISSN{0268-4012}
\href{https://doi.org/10.1016/j.ijinfomgt.2026.103031}{doi:\nolinkurl{10.1016/j.ijinfomgt.2026.103031}}


\bibitem[Belcak et~al\mbox{.}(2025)]%
        {belcak2025smalllanguagemodelsfuture}
\bibfield{author}{\bibinfo{person}{Peter Belcak}, \bibinfo{person}{Greg Heinrich}, \bibinfo{person}{Shizhe Diao}, \bibinfo{person}{Yonggan Fu}, \bibinfo{person}{Xin Dong}, \bibinfo{person}{Saurav Muralidharan}, \bibinfo{person}{Yingyan~Celine Lin}, {and} \bibinfo{person}{Pavlo Molchanov}.} \bibinfo{year}{2025}\natexlab{}.
\newblock \bibinfo{title}{Small Language Models are the Future of Agentic AI}.
\newblock
\showeprint[arxiv]{2506.02153}~[cs.AI]
\urldef\tempurl%
\url{https://arxiv.org/abs/2506.02153}
\showURL{%
\tempurl}


\bibitem[Cai et~al\mbox{.}(2024)]%
        {cai2024antagonisticai}
\bibfield{author}{\bibinfo{person}{Alice Cai}, \bibinfo{person}{Ian Arawjo}, {and} \bibinfo{person}{Elena~L. Glassman}.} \bibinfo{year}{2024}\natexlab{}.
\newblock \bibinfo{title}{Antagonistic AI}.
\newblock
\showeprint[arxiv]{2402.07350}~[cs.AI]
\urldef\tempurl%
\url{https://arxiv.org/abs/2402.07350}
\showURL{%
\tempurl}


\bibitem[Crisan et~al\mbox{.}(2022)]%
        {Crisan2022ModelCard}
\bibfield{author}{\bibinfo{person}{Anamaria Crisan}, \bibinfo{person}{Margaret Drouhard}, \bibinfo{person}{Jesse Vig}, {and} \bibinfo{person}{Nazneen Rajani}.} \bibinfo{year}{2022}\natexlab{}.
\newblock \showarticletitle{Interactive Model Cards: A Human-Centered Approach to Model Documentation}. In \bibinfo{booktitle}{\emph{Proceedings of the 2022 ACM Conference on Fairness, Accountability, and Transparency}} (Seoul, Republic of Korea) \emph{(\bibinfo{series}{FAccT '22})}. \bibinfo{publisher}{Association for Computing Machinery}, \bibinfo{address}{New York, NY, USA}, \bibinfo{pages}{427–439}.
\newblock
\showISBNx{9781450393522}
\href{https://doi.org/10.1145/3531146.3533108}{doi:\nolinkurl{10.1145/3531146.3533108}}


\bibitem[Das et~al\mbox{.}(2022)]%
        {das2022security}
\bibfield{author}{\bibinfo{person}{Sauvik Das}, \bibinfo{person}{Cori Faklaris}, \bibinfo{person}{Jason~I Hong}, \bibinfo{person}{Laura~A Dabbish}, {et~al\mbox{.}}} \bibinfo{year}{2022}\natexlab{}.
\newblock \showarticletitle{The security \& privacy acceptance framework (spaf)}.
\newblock \bibinfo{journal}{\emph{Foundations and Trends{\textregistered} in Privacy and Security}} \bibinfo{volume}{5}, \bibinfo{number}{1-2} (\bibinfo{year}{2022}), \bibinfo{pages}{1--143}.
\newblock


\bibitem[Hsieh et~al\mbox{.}(2023)]%
        {hsieh2023distillingstepbystepoutperforminglarger}
\bibfield{author}{\bibinfo{person}{Cheng-Yu Hsieh}, \bibinfo{person}{Chun-Liang Li}, \bibinfo{person}{Chih-Kuan Yeh}, \bibinfo{person}{Hootan Nakhost}, \bibinfo{person}{Yasuhisa Fujii}, \bibinfo{person}{Alexander Ratner}, \bibinfo{person}{Ranjay Krishna}, \bibinfo{person}{Chen-Yu Lee}, {and} \bibinfo{person}{Tomas Pfister}.} \bibinfo{year}{2023}\natexlab{}.
\newblock \bibinfo{title}{Distilling Step-by-Step! Outperforming Larger Language Models with Less Training Data and Smaller Model Sizes}.
\newblock
\showeprint[arxiv]{2305.02301}~[cs.CL]
\urldef\tempurl%
\url{https://arxiv.org/abs/2305.02301}
\showURL{%
\tempurl}


\bibitem[Kosmyna et~al\mbox{.}(2025)]%
        {kosmyna2025brainchatgptaccumulationcognitive}
\bibfield{author}{\bibinfo{person}{Nataliya Kosmyna}, \bibinfo{person}{Eugene Hauptmann}, \bibinfo{person}{Ye~Tong Yuan}, \bibinfo{person}{Jessica Situ}, \bibinfo{person}{Xian-Hao Liao}, \bibinfo{person}{Ashly~Vivian Beresnitzky}, \bibinfo{person}{Iris Braunstein}, {and} \bibinfo{person}{Pattie Maes}.} \bibinfo{year}{2025}\natexlab{}.
\newblock \bibinfo{title}{Your Brain on ChatGPT: Accumulation of Cognitive Debt when Using an AI Assistant for Essay Writing Task}.
\newblock
\showeprint[arxiv]{2506.08872}~[cs.AI]
\urldef\tempurl%
\url{https://arxiv.org/abs/2506.08872}
\showURL{%
\tempurl}


\bibitem[Laato et~al\mbox{.}(2022)]%
        {Laato2022ExplainAI}
\bibfield{author}{\bibinfo{person}{Samuli Laato}, \bibinfo{person}{Miika Tiainen}, \bibinfo{person}{A.K.M. {Najmul Islam}}, {and} \bibinfo{person}{Matti Mäntymäki}.} \bibinfo{year}{2022}\natexlab{}.
\newblock \showarticletitle{How to explain AI systems to end users: a systematic literature review and research agenda}.
\newblock \bibinfo{journal}{\emph{Internet Research}} \bibinfo{volume}{32}, \bibinfo{number}{7} (\bibinfo{year}{2022}), \bibinfo{pages}{1--31}.
\newblock
\showISSN{1066-2243}
\href{https://doi.org/10.1108/INTR-08-2021-0600}{doi:\nolinkurl{10.1108/INTR-08-2021-0600}}


\bibitem[Lee et~al\mbox{.}(2025)]%
        {Lee2025Thinking}
\bibfield{author}{\bibinfo{person}{Hao-Ping~(Hank) Lee}, \bibinfo{person}{Advait Sarkar}, \bibinfo{person}{Lev Tankelevitch}, \bibinfo{person}{Ian Drosos}, \bibinfo{person}{Sean Rintel}, \bibinfo{person}{Richard Banks}, {and} \bibinfo{person}{Nicholas Wilson}.} \bibinfo{year}{2025}\natexlab{}.
\newblock \showarticletitle{The Impact of Generative AI on Critical Thinking: Self-Reported Reductions in Cognitive Effort and Confidence Effects From a Survey of Knowledge Workers}. In \bibinfo{booktitle}{\emph{Proceedings of the 2025 CHI Conference on Human Factors in Computing Systems}} (Yokohama, Japan) \emph{(\bibinfo{series}{CHI '25})}. \bibinfo{publisher}{Association for Computing Machinery}, \bibinfo{address}{New York, NY, USA}, Article \bibinfo{articleno}{1121}, \bibinfo{numpages}{22}~pages.
\newblock
\showISBNx{9798400713941}
\href{https://doi.org/10.1145/3706598.3713778}{doi:\nolinkurl{10.1145/3706598.3713778}}


\bibitem[Lee et~al\mbox{.}(2026)]%
        {lee2026privyenvisioningmitigatingprivacy}
\bibfield{author}{\bibinfo{person}{Hao-Ping~(Hank) Lee}, \bibinfo{person}{Yu-Ju Yang}, \bibinfo{person}{Matthew Bilik}, \bibinfo{person}{Isadora Krsek}, \bibinfo{person}{Thomas~Serban von Davier}, \bibinfo{person}{Kyzyl Monteiro}, \bibinfo{person}{Jason Lin}, \bibinfo{person}{Shivani Agarwal}, \bibinfo{person}{Jodi Forlizzi}, {and} \bibinfo{person}{Sauvik Das}.} \bibinfo{year}{2026}\natexlab{}.
\newblock \showarticletitle{Privy: Envisioning and Mitigating Privacy Risks for Consumer-facing AI Product Concepts}. In \bibinfo{booktitle}{\emph{Proceedings of the 2026 {CHI} {Conference} on {Human} {Factors} in {Computing} {Systems}}}. \bibinfo{publisher}{Association for Computing Machinery}.
\newblock
\href{https://doi.org/10.1145/3772318.3791279}{doi:\nolinkurl{10.1145/3772318.3791279}}


\bibitem[Lee et~al\mbox{.}(2024)]%
        {Lee2024Taxonomy}
\bibfield{author}{\bibinfo{person}{Hao-Ping~(Hank) Lee}, \bibinfo{person}{Yu-Ju Yang}, \bibinfo{person}{Thomas~Serban Von~Davier}, \bibinfo{person}{Jodi Forlizzi}, {and} \bibinfo{person}{Sauvik Das}.} \bibinfo{year}{2024}\natexlab{}.
\newblock \showarticletitle{Deepfakes, Phrenology, Surveillance, and More! A Taxonomy of AI Privacy Risks}. In \bibinfo{booktitle}{\emph{Proceedings of the 2024 CHI Conference on Human Factors in Computing Systems}} (Honolulu, HI, USA) \emph{(\bibinfo{series}{CHI '24})}. \bibinfo{publisher}{Association for Computing Machinery}, \bibinfo{address}{New York, NY, USA}, Article \bibinfo{articleno}{775}, \bibinfo{numpages}{19}~pages.
\newblock
\showISBNx{9798400703300}
\href{https://doi.org/10.1145/3613904.3642116}{doi:\nolinkurl{10.1145/3613904.3642116}}


\bibitem[Lewis et~al\mbox{.}(2020)]%
        {Lewis2020RAG}
\bibfield{author}{\bibinfo{person}{Patrick Lewis}, \bibinfo{person}{Ethan Perez}, \bibinfo{person}{Aleksandra Piktus}, \bibinfo{person}{Fabio Petroni}, \bibinfo{person}{Vladimir Karpukhin}, \bibinfo{person}{Naman Goyal}, \bibinfo{person}{Heinrich K\"{u}ttler}, \bibinfo{person}{Mike Lewis}, \bibinfo{person}{Wen-tau Yih}, \bibinfo{person}{Tim Rockt\"{a}schel}, \bibinfo{person}{Sebastian Riedel}, {and} \bibinfo{person}{Douwe Kiela}.} \bibinfo{year}{2020}\natexlab{}.
\newblock \showarticletitle{Retrieval-Augmented Generation for Knowledge-Intensive NLP Tasks}. In \bibinfo{booktitle}{\emph{Advances in Neural Information Processing Systems}}, \bibfield{editor}{\bibinfo{person}{H.~Larochelle}, \bibinfo{person}{M.~Ranzato}, \bibinfo{person}{R.~Hadsell}, \bibinfo{person}{M.F. Balcan}, {and} \bibinfo{person}{H.~Lin}} (Eds.), Vol.~\bibinfo{volume}{33}. \bibinfo{publisher}{Curran Associates, Inc.}, \bibinfo{pages}{9459--9474}.
\newblock
\urldef\tempurl%
\url{https://proceedings.neurips.cc/paper_files/paper/2020/file/6b493230205f780e1bc26945df7481e5-Paper.pdf}
\showURL{%
\tempurl}


\bibitem[Liu et~al\mbox{.}(2025a)]%
        {liu2025interactingthoughtfulai}
\bibfield{author}{\bibinfo{person}{Xingyu~Bruce Liu}, \bibinfo{person}{Haijun Xia}, {and} \bibinfo{person}{Xiang~Anthony Chen}.} \bibinfo{year}{2025}\natexlab{a}.
\newblock \bibinfo{title}{Interacting with Thoughtful AI}.
\newblock
\showeprint[arxiv]{2502.18676}~[cs.HC]
\urldef\tempurl%
\url{https://arxiv.org/abs/2502.18676}
\showURL{%
\tempurl}


\bibitem[Liu et~al\mbox{.}(2025b)]%
        {liu2025harnessingcollaborativepowerlarge}
\bibfield{author}{\bibinfo{person}{Yang Liu}, \bibinfo{person}{Bingjie Yan}, \bibinfo{person}{Tianyuan Zou}, \bibinfo{person}{Jianqing Zhang}, \bibinfo{person}{Zixuan Gu}, \bibinfo{person}{Jianbing Ding}, \bibinfo{person}{Xidong Wang}, \bibinfo{person}{Jingyi Li}, \bibinfo{person}{Xiaozhou Ye}, \bibinfo{person}{Ye Ouyang}, \bibinfo{person}{Qiang Yang}, {and} \bibinfo{person}{Ya-Qin Zhang}.} \bibinfo{year}{2025}\natexlab{b}.
\newblock \bibinfo{title}{Towards Harnessing the Collaborative Power of Large and Small Models for Domain Tasks}.
\newblock
\showeprint[arxiv]{2504.17421}~[cs.LG]
\urldef\tempurl%
\url{https://arxiv.org/abs/2504.17421}
\showURL{%
\tempurl}


\bibitem[Meta(2024)]%
        {Llama3_1}
\bibfield{author}{\bibinfo{person}{Meta}.} \bibinfo{year}{2024}\natexlab{}.
\newblock \bibinfo{title}{Llama3.1}.
\newblock \bibinfo{howpublished}{\url{https://llama.meta.com/}}.
\newblock


\bibitem[Park et~al\mbox{.}(2024)]%
        {park2024thinkingassistantsllmbasedconversational}
\bibfield{author}{\bibinfo{person}{Soya Park}, \bibinfo{person}{Hari Subramonyam}, {and} \bibinfo{person}{Chinmay Kulkarni}.} \bibinfo{year}{2024}\natexlab{}.
\newblock \bibinfo{title}{Thinking Assistants: LLM-Based Conversational Assistants that Help Users Think By Asking rather than Answering}.
\newblock
\showeprint[arxiv]{2312.06024}~[cs.HC]
\urldef\tempurl%
\url{https://arxiv.org/abs/2312.06024}
\showURL{%
\tempurl}


\bibitem[Piorkowski et~al\mbox{.}(2020)]%
        {piorkowski2020evaluatingelicitinghighqualitydocumentation}
\bibfield{author}{\bibinfo{person}{David Piorkowski}, \bibinfo{person}{Daniel González}, \bibinfo{person}{John Richards}, {and} \bibinfo{person}{Stephanie Houde}.} \bibinfo{year}{2020}\natexlab{}.
\newblock \bibinfo{title}{Towards evaluating and eliciting high-quality documentation for intelligent systems}.
\newblock
\showeprint[arxiv]{2011.08774}~[cs.SE]
\urldef\tempurl%
\url{https://arxiv.org/abs/2011.08774}
\showURL{%
\tempurl}


\bibitem[Rezwana and Maher(2022)]%
        {Rezwana2022UserPerception}
\bibfield{author}{\bibinfo{person}{Jeba Rezwana} {and} \bibinfo{person}{Mary~Lou Maher}.} \bibinfo{year}{2022}\natexlab{}.
\newblock \showarticletitle{Understanding User Perceptions, Collaborative Experience and User Engagement in Different Human-AI Interaction Designs for Co-Creative Systems}. In \bibinfo{booktitle}{\emph{Proceedings of the 14th Conference on Creativity and Cognition}} (Venice, Italy) \emph{(\bibinfo{series}{C\&C '22})}. \bibinfo{publisher}{Association for Computing Machinery}, \bibinfo{address}{New York, NY, USA}, \bibinfo{pages}{38–48}.
\newblock
\showISBNx{9781450393270}
\href{https://doi.org/10.1145/3527927.3532789}{doi:\nolinkurl{10.1145/3527927.3532789}}


\bibitem[Salma et~al\mbox{.}(2025)]%
        {Salma2025Paradox}
\bibfield{author}{\bibinfo{person}{Zainab Salma}, \bibinfo{person}{Raquel Hijón-Neira}, {and} \bibinfo{person}{Celeste Pizarro}.} \bibinfo{year}{2025}\natexlab{}.
\newblock \showarticletitle{Designing Co-Creative Systems: Five Paradoxes in Human–AI Collaboration}.
\newblock \bibinfo{journal}{\emph{Information}} \bibinfo{volume}{16}, \bibinfo{number}{10} (\bibinfo{year}{2025}).
\newblock
\showISSN{2078-2489}
\href{https://doi.org/10.3390/info16100909}{doi:\nolinkurl{10.3390/info16100909}}


\bibitem[Sarkar et~al\mbox{.}(2024)]%
        {sarkar2024copilotautopilotgenerativeais}
\bibfield{author}{\bibinfo{person}{Advait Sarkar}, \bibinfo{person}{Xiaotong}, \bibinfo{person}{Xu}, \bibinfo{person}{Neil Toronto}, \bibinfo{person}{Ian Drosos}, {and} \bibinfo{person}{Christian Poelitz}.} \bibinfo{year}{2024}\natexlab{}.
\newblock \bibinfo{title}{When Copilot Becomes Autopilot: Generative AI's Critical Risk to Knowledge Work and a Critical Solution}.
\newblock
\showeprint[arxiv]{2412.15030}~[cs.HC]
\urldef\tempurl%
\url{https://arxiv.org/abs/2412.15030}
\showURL{%
\tempurl}


\bibitem[Skidmore and Murakami(2016)]%
        {Skidmore2016}
\bibfield{author}{\bibinfo{person}{David Skidmore} {and} \bibinfo{person}{Kyoko Murakami}.} \bibinfo{year}{2016}\natexlab{}.
\newblock \bibinfo{booktitle}{\emph{Dialogic Pedagogy: An Introduction}}.
\newblock \bibinfo{publisher}{Multilingual Matters}, \bibinfo{pages}{1--16}.
\newblock
\showISBNx{9781783098408}
\urldef\tempurl%
\url{http://ebookcentral.proquest.com/lib/bath/detail.action?docID=4614619.}
\showURL{%
\tempurl}


\bibitem[von Davier et~al\mbox{.}(2025)]%
        {vonDavier2025ArtBot}
\bibfield{author}{\bibinfo{person}{Thomas~Serban von Davier}, \bibinfo{person}{Aaron John~Henry Larsen}, \bibinfo{person}{Max Van~Kleek}, {and} \bibinfo{person}{Nigel Shadbolt}.} \bibinfo{year}{2025}\natexlab{}.
\newblock \showarticletitle{ArtBot: An Exploration into AI’s Potential for Guiding Art Analysis}. In \bibinfo{booktitle}{\emph{Proceedings of the Extended Abstracts of the CHI Conference on Human Factors in Computing Systems}} (Yokohama, Japan) \emph{(\bibinfo{series}{CHI EA '25})}. \bibinfo{publisher}{Association for Computing Machinery}, \bibinfo{address}{New York, NY, USA}, Article \bibinfo{articleno}{77}, \bibinfo{numpages}{11}~pages.
\newblock
\showISBNx{9798400713958}
\href{https://doi.org/10.1145/3706599.3720181}{doi:\nolinkurl{10.1145/3706599.3720181}}


\bibitem[Wang et~al\mbox{.}(2025)]%
        {Wang2025MentalModels}
\bibfield{author}{\bibinfo{person}{Xingyi Wang}, \bibinfo{person}{Xiaozheng Wang}, \bibinfo{person}{Sunyup Park}, {and} \bibinfo{person}{Yaxing Yao}.} \bibinfo{year}{2025}\natexlab{}.
\newblock \showarticletitle{Mental Models of Generative AI Chatbot Ecosystems}. In \bibinfo{booktitle}{\emph{Proceedings of the 30th International Conference on Intelligent User Interfaces}} \emph{(\bibinfo{series}{IUI '25})}. \bibinfo{publisher}{Association for Computing Machinery}, \bibinfo{address}{New York, NY, USA}, \bibinfo{pages}{1016–1031}.
\newblock
\showISBNx{9798400713064}
\href{https://doi.org/10.1145/3708359.3712125}{doi:\nolinkurl{10.1145/3708359.3712125}}


\bibitem[Wang et~al\mbox{.}(2024)]%
        {wang2024FarSight}
\bibfield{author}{\bibinfo{person}{Zijie~J. Wang}, \bibinfo{person}{Chinmay Kulkarni}, \bibinfo{person}{Lauren Wilcox}, \bibinfo{person}{Michael Terry}, {and} \bibinfo{person}{Michael Madaio}.} \bibinfo{year}{2024}\natexlab{}.
\newblock \showarticletitle{Farsight: Fostering Responsible AI Awareness During AI Application Prototyping}. In \bibinfo{booktitle}{\emph{Proceedings of the 2024 CHI Conference on Human Factors in Computing Systems}} (Honolulu, HI, USA) \emph{(\bibinfo{series}{CHI '24})}. \bibinfo{publisher}{Association for Computing Machinery}, \bibinfo{address}{New York, NY, USA}, Article \bibinfo{articleno}{976}, \bibinfo{numpages}{40}~pages.
\newblock
\showISBNx{9798400703300}
\href{https://doi.org/10.1145/3613904.3642335}{doi:\nolinkurl{10.1145/3613904.3642335}}


\bibitem[Wei et~al\mbox{.}(2022)]%
        {wei2022chain}
\bibfield{author}{\bibinfo{person}{Jason Wei}, \bibinfo{person}{Xuezhi Wang}, \bibinfo{person}{Dale Schuurmans}, \bibinfo{person}{Maarten Bosma}, \bibinfo{person}{Fei Xia}, \bibinfo{person}{Ed Chi}, \bibinfo{person}{Quoc~V Le}, \bibinfo{person}{Denny Zhou}, {et~al\mbox{.}}} \bibinfo{year}{2022}\natexlab{}.
\newblock \showarticletitle{Chain-of-thought prompting elicits reasoning in large language models}.
\newblock \bibinfo{journal}{\emph{Advances in neural information processing systems}}  \bibinfo{volume}{35} (\bibinfo{year}{2022}), \bibinfo{pages}{24824--24837}.
\newblock


\bibitem[Wiberg and Stolterman~Bergqvist(2023)]%
        {Wiberg2023Automation}
\bibfield{author}{\bibinfo{person}{Mikael Wiberg} {and} \bibinfo{person}{Erik Stolterman~Bergqvist}.} \bibinfo{year}{2023}\natexlab{}.
\newblock \showarticletitle{Automation of interaction—interaction design at the crossroads of user experience (UX) and artificial intelligence (AI)}.
\newblock \bibinfo{journal}{\emph{Personal and Ubiquitous Computing}}  \bibinfo{volume}{27} (\bibinfo{year}{2023}), \bibinfo{pages}{2281--2290}.
\newblock
\href{https://doi.org/10.1007/s00779-023-01779-0}{doi:\nolinkurl{10.1007/s00779-023-01779-0}}


\bibitem[Ye et~al\mbox{.}(2024)]%
        {ye2024languagemodelscriticalthinking}
\bibfield{author}{\bibinfo{person}{Andre Ye}, \bibinfo{person}{Jared Moore}, \bibinfo{person}{Rose Novick}, {and} \bibinfo{person}{Amy~X. Zhang}.} \bibinfo{year}{2024}\natexlab{}.
\newblock \bibinfo{title}{Language Models as Critical Thinking Tools: A Case Study of Philosophers}.
\newblock
\showeprint[arxiv]{2404.04516}~[cs.HC]
\urldef\tempurl%
\url{https://arxiv.org/abs/2404.04516}
\showURL{%
\tempurl}


\end{thebibliography}

\appendix
\onecolumn
\section{System Screenshots}

\end{document}